\documentclass[journal=nalefd,manuscript=article,layout=twocolumn]{achemso}

\usepackage{graphicx}
\usepackage[svgnames]{xcolor}
\usepackage{mathtools}
\usepackage{hyperref}
\hypersetup{pdftitle={GrhBN},
pdfauthor={ICN2},
pdfnewwindow=true, 
colorlinks=true,   
linkcolor=DarkBlue,     
citecolor=DarkBlue,   
filecolor=DarkBlue, 
urlcolor=DarkBlue      
}
\usepackage{achemso}
\usepackage{natbib}

\setkeys{acs}{super=false}
\setkeys{acs}{etalmode=truncate}
\setkeys{acs}{hyperref=true}
\setkeys{acs}{maxauthors=4}

\DeclareMathOperator{\Tr}{Tr}

\author{J.E. Barrios Vargas}
\affiliation{Catalan Institute of Nanoscience and Nanotechnology (ICN2), CSIC and The Barcelona Institute of Science and Technology, Campus UAB, 08193 Barcelona, Spain}
\email{jose.barrios@icn2.cat}
\author{B. Mortazavi}
\affiliation{Institute of Structural Mechanics, Bauhaus-Universit\"at Weimar, Marienstr. 15, D-99423 Weimar, Germany}
\author{A.W. Cummings} 
\affiliation{Catalan Institute of Nanoscience and Nanotechnology (ICN2), CSIC and The Barcelona Institute of Science and Technology, Campus UAB, 08193 Barcelona, Spain}
\author{R. Martinez-Gordillo}
\affiliation{CINaM - Centre Interdisciplinaire de Nanoscience de Marseille, France}
\author{M. Pruneda}
\affiliation{Catalan Institute of Nanoscience and Nanotechnology (ICN2), CSIC and The Barcelona Institute of Science and Technology, Campus UAB, 08193 Barcelona, Spain}
\author{L. Colombo}
\affiliation{Catalan Institute of Nanoscience and Nanotechnology (ICN2), CSIC and The Barcelona Institute of Science and Technology, Campus UAB, 08193 Barcelona, Spain}
\alsoaffiliation{Dipartimento di Fisica, Università di Cagliari, Cittadella Universitaria, I-09042 Monserrato (Ca), Italy}
\author{T. Rabczuk} 
\affiliation{Institute of Structural Mechanics, Bauhaus-Universit\"at Weimar, Marienstr. 15, D-99423 Weimar, Germany}
\author{S. Roche}
\email{stephan.roche@icn2.cat}
\affiliation{Catalan Institute of Nanoscience and Nanotechnology (ICN2), CSIC and The Barcelona Institute of Science and Technology, Campus UAB, 08193 Barcelona, Spain}
\alsoaffiliation{ICREA - Instituci\'{o} Catalana de Recerca i Estudis Avan\c{c}ats, 08010 Barcelona, Spain}

\title{Electrical and thermal transport in coplanar polycrystalline graphene-hBN heterostructures}

\begin{document}

\begin{abstract}
We present a theoretical study of electronic and thermal transport in polycrystalline heterostructures combining graphene (G) and hexagonal boron nitride (hBN) grains of varying size and distribution. By increasing the hBN grain density from a few percent to $100\%$, the system evolves from a good conductor to an insulator, with the mobility dropping by orders of magnitude and the sheet resistance reaching the M$\Omega$ regime. The Seebeck coefficient is suppressed above $40\%$ mixing, while the thermal conductivity of polycrystalline hBN is found to be on the order of $30-120\,{\rm W}{\rm m}^{-1}{\rm K}^{-1}$. These results, agreeing with available experimental data, provide guidelines for tuning G-hBN properties in the context of two-dimensional materials engineering. In particular, while we proved that both electrical and thermal properties are largely affected by morphological features (like e.g. by the grain size and composition), we find in all cases that nm-sized polycrystalline G-hBN heterostructures are not good thermoelectric materials.
\end{abstract}

{\color{Blue}
{{\bf Keywords:} Polycrystalline graphene, boron nitride, chemical vapor deposition, grain boundary, electrical properties, thermal properties, thermoelectrics.}\\
}

{\bf Introduction.}
Owing to a small lattice mismatch (2\%), graphene and hexagonal boron nitride can be assembled in coplanar two-dimensional heterostructures~\cite{Rubio2010}. Such atomic sheets, covering a wide range of compositions, result in new materials with properties complementary to those of graphene and hBN, such as tunable bandgap optoelectronic materials~\cite{Ci2010}. Graphene is well appreciated for its high electrical~\cite{Novoselov2004} and thermal conductivities~\cite{Baladin2008}, whereas hBN is an electrical insulator with to date an unmeasured thermal conductivity~\cite{Jo2013, Wang2016}. Large-scale coplanar G-hBN heterostructures have been successfully fabricated using chemical vapor deposition (CVD), enabling the possible control of periodic arrangements of domains whose sizes range from tens of nanometers to millimeters~\cite{Levendorf2012, Liu2013, Gang2013, Liu2014}. Their charge transport properties can be, however, quite surprising, such as the presence of a metal-insulator transition~\cite{Zhao2012,Lars2012,Gong2014} and anomalous transport phenomena, that is not fully understood.~\cite{Li2012} Additionally, fast CVD growth results in polycrystalline materials with grains of varying sizes and morphologies, and the electronic and thermal properties of these materials are limited by the presence of grain boundaries (GBs)~\cite{Tuan2013, Cummings2014, Mortazavi2014, Hahn2016,Andreas2017}.

In polycrystalline graphene, GBs are characterized by Van Hove singularities near the Dirac point~\cite{Ma2014, Yann2014, Adina2016}, whereas in hBN the GBs reduce the bandgap and introduce gap states generated by the presence of B-B or N-N bonds~\cite{Qiucheng2015}. The interface between G and hBN is also expected to give rise to local boundary states, especially at low energies~\cite{Robert2014, Jiong2014}. GBs are also usually accompanied by local structural deformation, which enhances phonon scattering and thus lowers thermal conduction. The thermal properties of polycrystalline graphene have been theoretically calculated using molecular dynamics simulations as a function of average grain size~\cite{Wang2014, Mortazavi2014, Liu2014Th, Hahn2016}, in fair agreement with experimental results~\cite{Baladin2008}.

Recently, a sample of CVD-grown graphene was gradually converted into hBN, and it was observed that chemical substitutions are initiated around structural defects. This process of conversion demonstrated a fine tunability between highly conductive graphene and insulating hBN~\cite{Gong2014}. To date however, the electronic and thermal properties of CVD-grown hybridized G-hBN heterostructures are poorly understood, and their potential use in energy harvesting, optoelectronic, or nanoelectronic applications remains unclear.

Here we use quantum transport and molecular dynamics (MD) simulations to calculate the electronic and thermal properties of polycrystalline G-hBN heterostructures with varying grain size and distribution. The electronic mobility and sheet resistance are studied as a function of the density of hBN grains, which ranges from a few percent to full coverage. The contribution of GB interface states to the transport properties is also illustrated and quantified. By performing a complete calculation of thermal and electrical transport, we estimate the thermoelectric conversion ratio and find that it remains far too low to be useful for energy harvesting applications.

\begin{figure}[h!]
\includegraphics[width=1.0\linewidth]{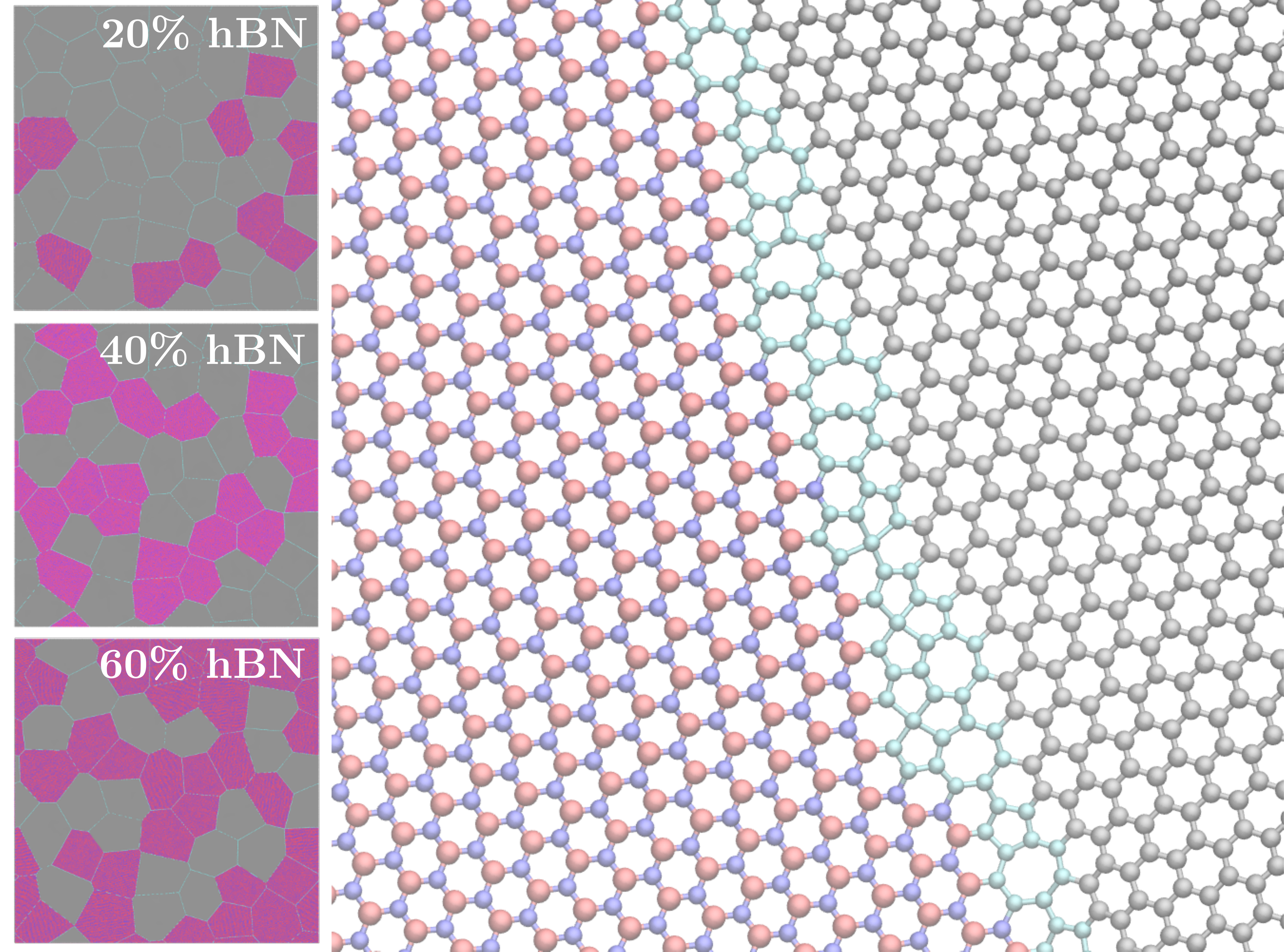}
\caption{Left panel: square periodic polycrystalline structures with three different concentrations of hBN (20\%, 40\% and 60\%). Right panel: magnification of the polycrystalline structure showing a typical interface between graphene and hBN grains.}
\label{Samples}
\end{figure}

{\bf Generation of samples.}
Polycrystalline G-hBN heterostructures with uniform average grain size were generated using a Voronoi algorithm, resulting in large square periodic samples containing up to 3 million atoms~\cite{Mortazavi2014, Mortazavi2015}. The algorithm starts with a random selection of nucleation centers within a square cell of predefined dimension, which dictates the average grain size as $L_{\rm grain} = \sqrt{L^2/n_{\rm grains}}$, where $L$ is the sample length and $n_{\rm grains}$ is the number of grains. Next we set a random crystal orientation for each nucleation site and we use a Voronoi method to construct the grains. The atoms along the GBs with separation below 0.1~nm are removed, and an MD annealing process is used to construct the GBs, setting all the atoms as carbon. We use the LAMMPS simulation package~\cite{Plimpton1995}, the second-generation reactive empirical bond order potential~\cite{Brenner2002}, and a small time increment of 0.1~fs. The annealing starts with a 3-ps equilibration at room temperature using the Nos\'e-Hoover thermostat, continues with a heating up to 3000~K for 12~ps and keeping this temperature for 3~ps, and ends with a cooling back down to room temperature for 10~ps. Finally, based the concentration of hBN, we assign which grains are graphene and which ones are hBN (Figure~\ref{Samples}).\\

{\bf Electronic properties.}
We describe the electronic properties of the G-hBN heterostructures with a tight-binding Hamiltonian
\begin{align}
\mathcal{H} = \sum_{\mathbf{r}_i} \varepsilon_i(\mathbf{r}_i) |\mathbf{r}_i \rangle \langle \mathbf{r}_i| 
+ \sum_{\langle \mathbf{r}_i,\mathbf{r}_j \rangle} t_{i,j} |\mathbf{r}_i \rangle \langle \mathbf{r}_j |\,,
\end{align}
where $\varepsilon_i(\mathbf{r}_i)$ is the on-site potential of each atom and $t_{i,j}$ is the hopping between nearest neighbors. In systems containing 1D interfaces between two different 2D materials, the electronic properties are sensitive to the interface termination, and thus care must be taken when describing the GBs between graphene and hBN grains. For example, zigzag BN nanoribbons are polar, presenting bound charge of opposite signs at the B and N edges. In hybrid systems, mobile electrons from the graphene will tend to screen the excess interfacial charge, which changes the potential profile across the GB. Therefore, we modify the on-site term of the Hamiltonian to include a position-dependent electrostatic potential, which can be derived from the screened Poisson equation considering point charges midway between the C-B or C-N interfacial bonds. The on-site term of the TB Hamiltonian can then be written as~\cite{Rafael2014thesis}
\begin{align}\label{eq:onsite}
\varepsilon_{i}(\mathbf{r}_i) = \varepsilon_{i0}
+\sum_\alpha^{n_q} \frac{A_i^\mathrm{B}}{|\mathbf{r}_i-\mathbf{r}_\alpha^\mathrm{B}|} 
e^{-\frac{|\mathbf{r}_i-\mathbf{r}_\alpha^\mathrm{B}|}{\lambda_i}} \nonumber \\
-\sum_\alpha^{n_q} \frac{A_i^\mathrm{N}}{|\mathbf{r}_i-\mathbf{r}_\alpha^\mathrm{N}|} 
e^{-\frac{|\mathbf{r}_i-\mathbf{r}_\alpha^\mathrm{N}|}{\lambda_i}},
\end{align}
where $\varepsilon_{i}(\mathbf{r}_i)$ denotes the on-site energy for an atom of type $i$ (either carbon, boron, or nitrogen) at position $\mathbf{r}_i$, $\varepsilon_{i0}$ is the on-site energy of atoms far from the GBs, $A_i^\mathrm{B}$ ($A_i^\mathrm{N}$) is the strength of the potential arising from the C-B (C-N) interface, $\mathbf{r}_\alpha^\mathrm{B}$ ($\mathbf{r}_\alpha^\mathrm{N}$) is the position of the excess charge at the C-B (C-N) interface, $\lambda_i$ is the decay length of the interface potential, and the sum is done for all $n_q$ charges within a radius of $1\,{\rm nm}$. The onsite potential and nearest-neighbor hopping parameters have been derived from a Wannierization of DFT calculations and are given in Table~\ref{TabTBH} (see Supplementary Information for more details). Finally, because the GBs contain non-hexagonal rings, B-B or N-N bonds will be present. For these bonds we set $t_{\rm BN}$ as the hopping parameter, while the on-site energy is taken as $1.1\varepsilon_{i0}$.

\begin{table}[h!]
\begin{tabular}{cccc}
\hline
\multicolumn{4}{l}{\bf On-site energy (eV)}\\
\hline
$\varepsilon_{\rm C0}$ & $\varepsilon_{\rm B0}$ & $\varepsilon_{\rm N0}$& \\
0.0& 3.09&-1.89\\
\hline
\multicolumn{4}{l}{\bf Boundary Electrostatic Potential parameters}\\
\hline
$\lambda_{\rm C}$ & $\lambda_{\rm B}=\lambda_{\rm N}$ & $A_{i}^{\rm B}=A_{i}^{\rm N}$ & \\
6.78 \AA &12.56 \AA&0.56 eV$\cdot$\AA&\\
\hline
\multicolumn{4}{l}{\bf Nearest-neighbor hoppings (eV)}\\
\hline
$t_{\rm CC}$&$t_{\rm CB}$&$t_{\rm CN}$&$t_{\rm BN}$\\
-2.99& -2.68& -2.79& -3.03\\
\end{tabular}
\caption{On-site and nearest-neighbor tight-binding Hamiltonian parameters.}
\label{TabTBH}
\end{table}

\begin{figure}[h!]
\begin{tabular}{c}
\includegraphics[width=0.85\linewidth]{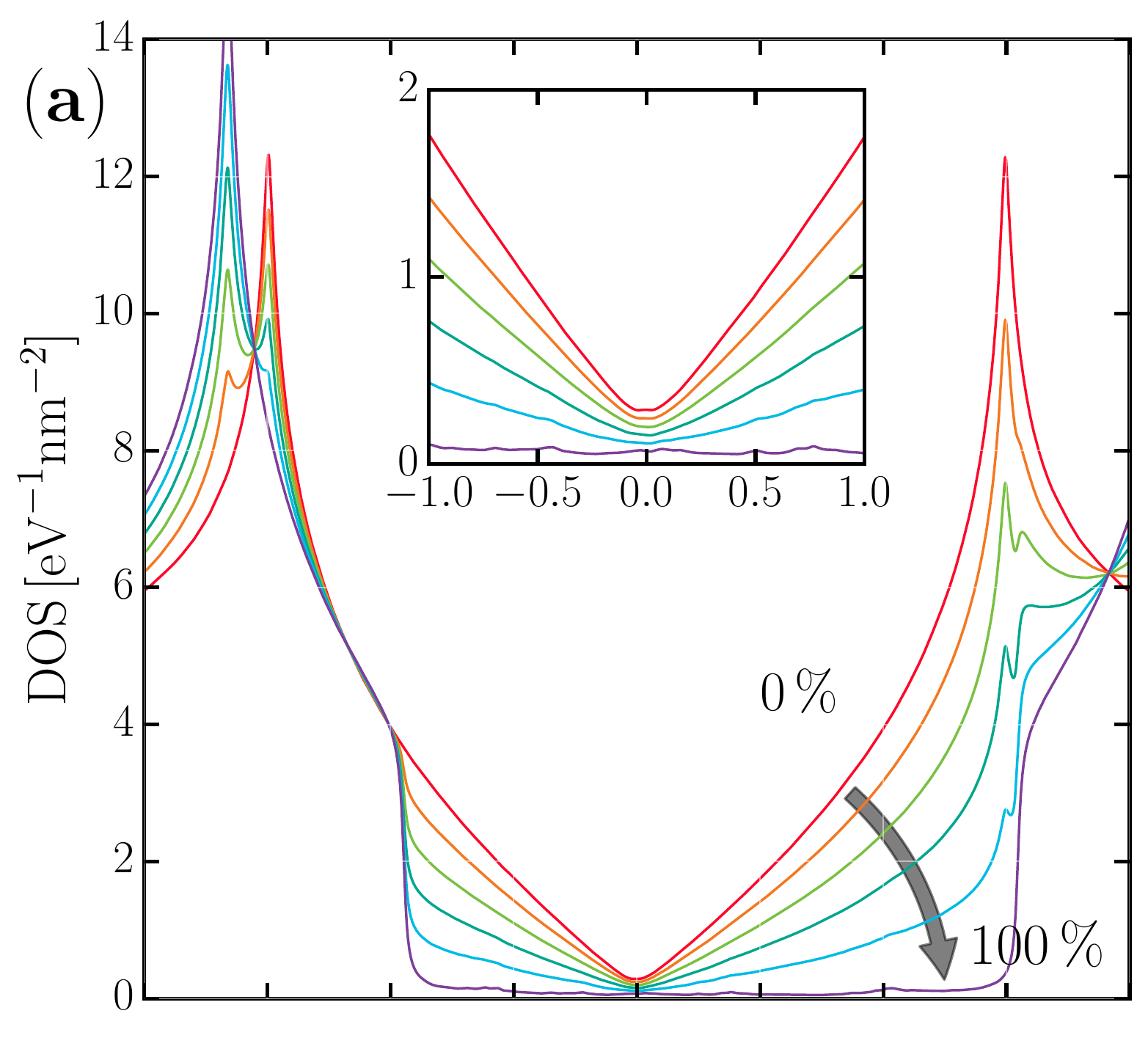}\\
\includegraphics[width=0.85\linewidth]{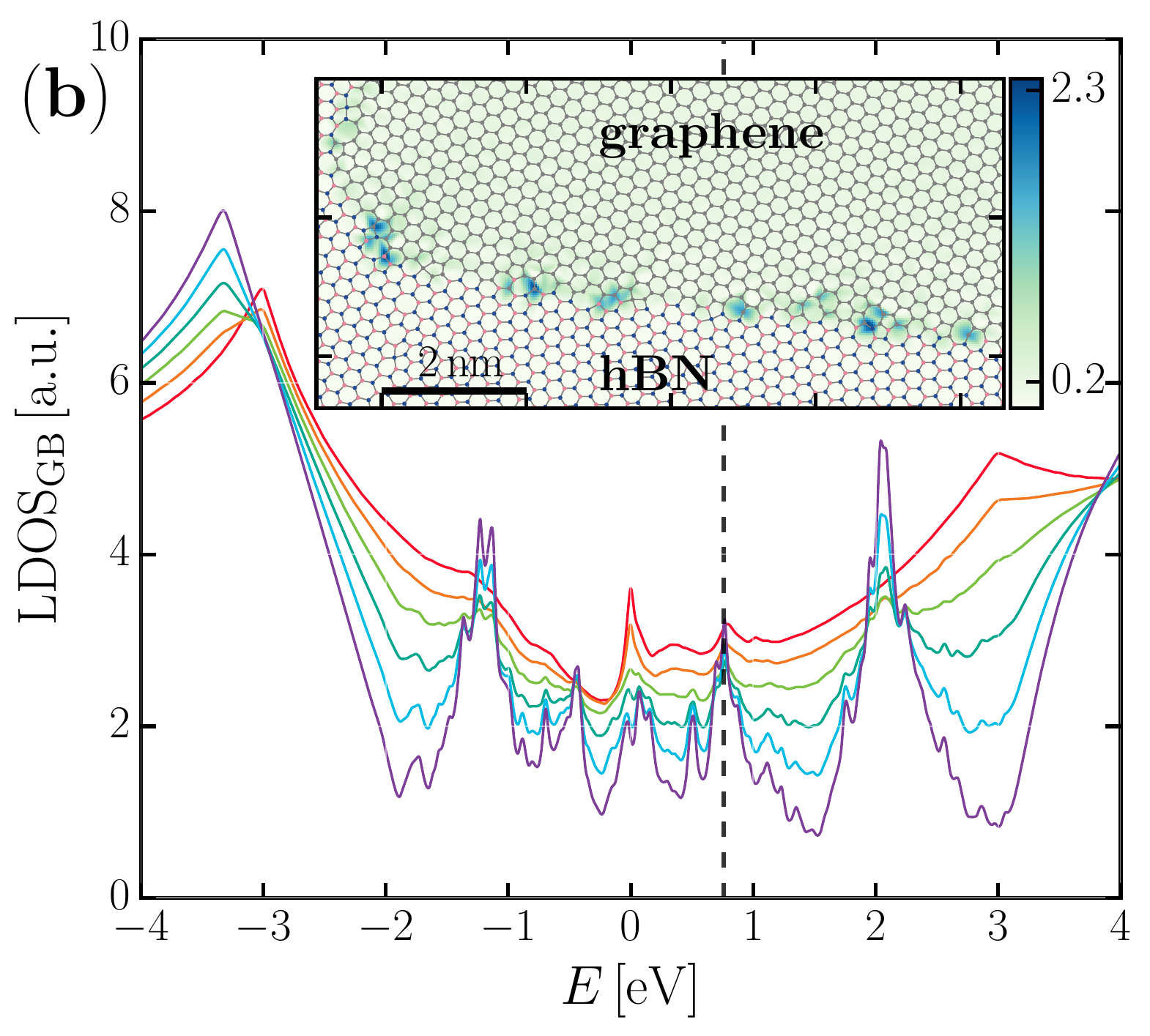}
\end{tabular}
\caption{(a) DOS of the polycrystalline lattice with increasing hBN grain density, in steps of 20\%, for an average grain size of 40 nm. Inset: magnification of the DOS in the interval [-1,1] eV. (b) ${\rm LDOS}_{\rm GB}$ for the same parameters. Inset: ${\rm LDOS}_{\rm GB}$ projected around a G-hBN interface, corresponding to the energy marked with the dashed line in the main frame.}
\label{DOS}
\end{figure}

We calculate the electronic density of states (DOS) using the Lanczos recursion method with an energy resolution of $\eta=kT=26\,{\rm meV}$ ($T=300\,{\rm K}$). Figure~\ref{DOS}(a) shows the DOS with increasing hBN grain density in steps of 20\%, for an average grain size of 40 nm. The gap is seen to progressively widen with increasing hBN concentration, but with a faster decay on the electron side of the spectrum. This electron-hole asymmetry stems from the GB states, which generate more resonances on the electron side. This can be seen more clearly for 100\% hBN, where the formation of boundary states, with energy lying inside the gap, is illustrated by the local density of states projected over all the GB sites (${\rm LDOS}_{\rm GB}$; Figure~\ref{DOS}(b)). The energy resonances at -1.2 and 2 eV has been observed experimentally, which can be associated to homoelemental bonds in the GB.\cite{Qiucheng2015} Besides, we observe other peaks at $E=0.0$ and $0.76$ eV, both are found for polycrystalline graphene and hBN, which suggest specific fingerprints of the structural morphology of grain boundaries. These states are mainly localized at the GBs, as visualized in the inset of Figure~\ref{DOS}(b), with stronger energy resonances on the electron side of the spectrum (see additional ${\rm LDOS}_{\rm GB}$ projected around a G-hBN interface in Supplementary Information). The presence of such states could be at the origin of the finite electrical conductivity computed for polycrystalline hBN (see below).

We next evaluate the electronic transport properties using a real-space order-N wave packet propagation method~\cite{Roche1999, RocheBook}. The core of this method is to calculate the time-dependent diffusion coefficient as
\begin{equation}
D(E,t) = \frac{\partial}{\partial t}\Delta X^2(E,t),
\label{diffusion}
\end{equation}
where $\Delta X^2$ is the mean-square displacement of the wave packet
\begin{equation}
\Delta X^2(E,t) = \frac{\Tr[\delta (E-\hat{H}) |\hat{X}(t) - \hat{X}(0)|^2]}{\rho(E)},
\label{msd}
\end{equation}
and $\rho(E) = \Tr[\delta (E-\hat{H})]$ is the DOS. We evaluate the trace using the Lanczos recursion and the same parameters as the DOS. We calculate the energy-dependent semiclassical conductivity as $\sigma(E) = e^2 \rho(E) \tilde{D}(E)$, where $\tilde{D}(E)$ is the value of the diffusion coefficient when the mean displacement has reached six times the average grain size (see Supplementary Information).

In Figure~\ref{sigma}(a) we report $\sigma(E)$, where a drop of more than two orders of magnitude is observed near the charge neutrality point with increasing hBN concentration. To further clarify the impact of the density of hBN grains, we fix the carrier concentration to $n = 0.3 \times 10^{12}\,{\rm cm}^{-2}$, which is a typical value for graphene on SiO$_2$\cite{Dorgan2010}, and evaluate the charge mobility $\mu = \sigma(n) / n$, shown in Figure~\ref{sigma}(b). The sheet resistance $R$ is shown in the inset of Figure~\ref{sigma}(b), where one can see that the maximum value for 100\% hBN is 
about $1\,{\rm M\Omega}$. Experimentally, a sheet resistance of a few ${\rm G\Omega}$ has been measured,~\cite{Gong2014}; this value is consistent with our calculations as a consecuence of the grain size scaling of the electrical properties.~\cite{Andreas2017} Additionally, we estimate the GB-resistivity, $\rho_{\rm GB}$, using an ohmic scaling analysis~\cite{Cummings2014,Andreas2017},
\begin{align}
R = R^{0} +\frac{\rho_{\rm GB}}{L_{\rm grain}}\,,
\end{align}
where $R$ and $R^{0}$ are the sheet resistances of the polycrystalline sample and the individual grains, respectively. The estimated resistivity for the G-G interface is $0.12$ and for hBN-hBN is 5.93~${\rm k\Omega}\cdot \mu{\rm m}$ (see Supplementary Information).

To complement the information about the electronic properties, we evaluate the Seebeck coefficient
\begin{align}
S(E) = -\frac{1}{|e|T} \frac{\displaystyle \int\displaylimits_{-\infty}^{\infty} (E'-E)G(E')\bigg(-\frac{\partial f}{\partial E'}\bigg) dE' }
{\displaystyle \int\displaylimits_{-\infty}^{\infty} G(E')\bigg(-\frac{\partial f}{\partial E'}\bigg) dE'}\,,
\end{align}
where $G$ is the sheet conductance and $f$ is the Fermi distribution. As shown in Figure~\ref{sigma}(c), the Seebeck coefficient of the polycrystalline samples is reduced compared to pristine graphene~\cite{Woessner2016}, but is insensitive to hBN concentrations below 40\%. However, beyond 40\% the thermoelectric capability is strongly suppressed. \\

\begin{figure}[h!]

\includegraphics[width=0.95\linewidth]{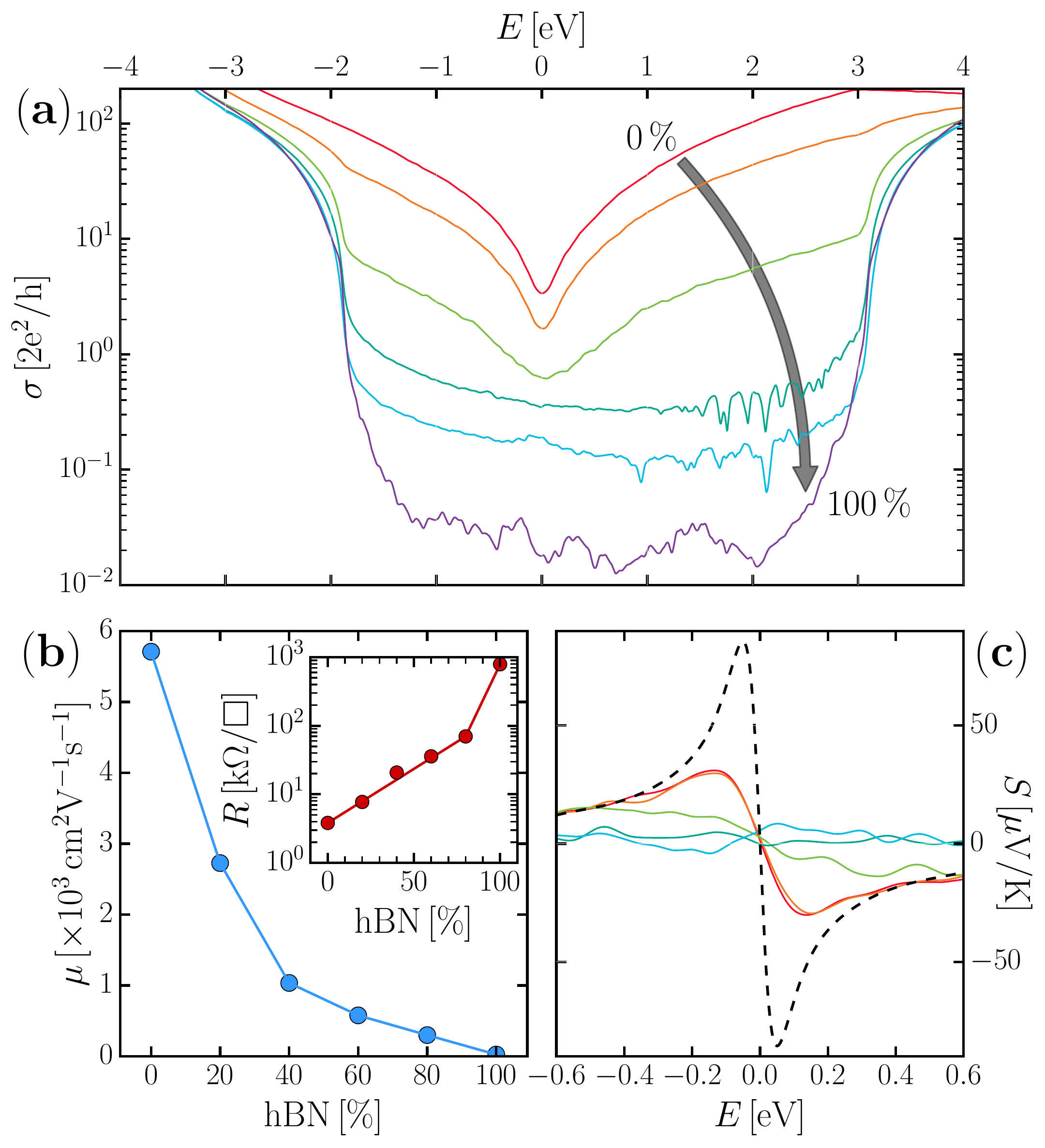}

\caption{(a) Conductivity versus energy for various hBN grain densities, with an average grain size of 40 nm. (b) Mobility as a function of hBN concentration for a fixed carrier density $n = 0.3 \times 10^{12}\,{\rm cm}^{-2}$. Inset: the sheet resistance for the same carrier density. (c) Seebeck coefficient with increasing hBN grain density, with the dashed line showing the pristine graphene value.}
\label{sigma}
\end{figure}

\begin{figure}[h!]
\includegraphics[width=1.0\linewidth]{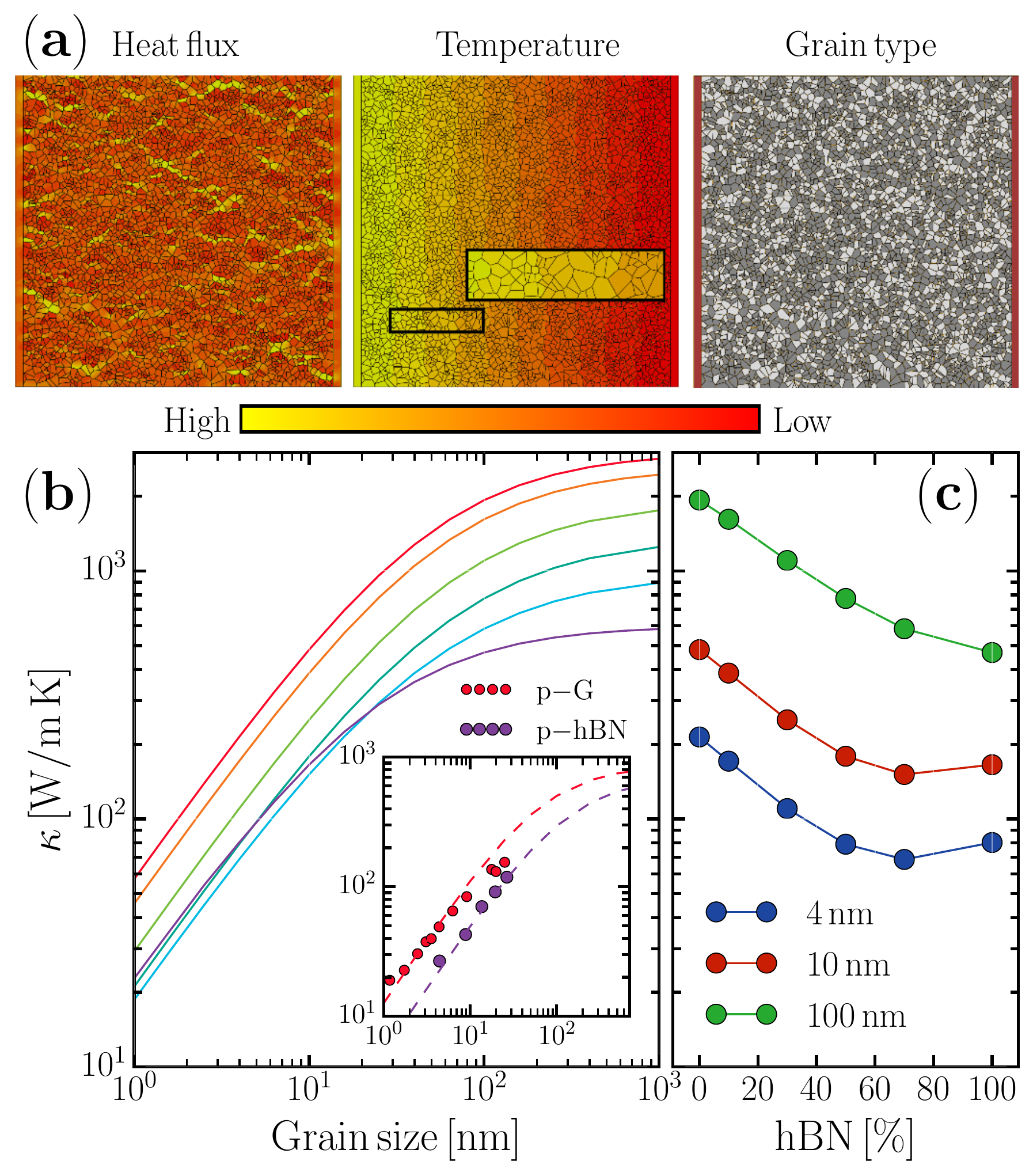}
\caption{(a) Heat flux (left panel) and temperature profile (center panel) calculated with FE for the granular mesh shown in the right panel. (b) Thermal conductivity as a function of the average grain size calculated with FE. Inset: symbols show the polycrystalline graphene (p-G) and hBN (p-hBN) thermal conductivities calculated using AEMD, while the dashed lines show the extrapolated scaling behavior using the extracted GB conductance. (c) Thermal conductivity as a function of the hBN grain density for different average grain sizes using the FE method.}
\label{thermal}
\end{figure}

{\bf Thermal properties.} In order to evaluate the thermal conductivity as a function of the grain size, we construct a finite element (FE) model in the ABAQUS package with 4000 grains constructed as Voronoi cells (right panel Figure~\ref{thermal}(a)). Using six representative pentagon-heptagon GB structures, we extract the GB thermal conductance for G-G, G-hBN and hBN-hBN interfaces by performing a non-equilibrium molecular dynamics (NEMD) calculation with LAMMPS (see Supplementary Information for details); which are introduced as contact conductances between interfaces. In the FE model, we include two highly conductive strips at the two ends of the structure\cite{Mortazavi2014, Mortazavi2015} and fix the ingoing (outgoing) heat flux on the left (right) side, $h_{\rm f}$. Then, we evaluate the steady-state temperature profile along the sample and use the $\Delta T$ between the strips to evaluate the effective thermal conductivity of the sample as
\begin{align}
\kappa = h_{\rm f}\frac{L}{\Delta T}\,,
\end{align}
where $L$ is the sample length. We calculate the themal conductivity for 16 grain sizes between 1-1000 nm while changing the concentration of hBN (Figure~\ref{thermal}(b)). The scaling of $\kappa$ shows that the impact of the GBs on thermal transport becomes negligible for grain sizes above 100 nm, which suggest that heat carriers with mean free path longer than 100 nm bring low contribution to $\kappa$. Figure~\ref{thermal}(c) displays the thermal conductivity as a function of the hBN grain density where we observe that, for small average grain size, the minimum of thermal conductivity occurs near 70\% hBN, similar to prior estimates~\cite{Sevincli2011}. This minimum can be rationalized by the fact that the thermal conductance for the G-hBN interface is lower than that of the hBN-hBN and G-G interfaces. For larger grain sizes, where the GBs no longer dominate the thermal transport, we observe a monotonic scaling of $\kappa$ with hBN grain density, as the thermal conductivity of pristine hBN is lower than that of pristine graphene.

In order to validate the above FE analysis we perform an independent investigation based on MD simulations. The goal is to provide evidence that the FE analysis, although missing most of the atomic-scale details, nevertheless provides the correct gross features on thermal transport across hBN and graphene GBs. The fully atomistic study of the thermal conductivity employs an approach-to-equilibrium molecular dynamics (AEMD) method following the same approach as in Ref.~\cite{Hahn2016} using the Tersoff BNC potential~\cite{Kinaci2012} (see Supplementary Information). While NEMD provides direct access to the temperature drop across the GB, which is the relevant quantity needed to calculate the interface thermal resistance (and, therefore, the GB conductance), the AEMD approach is better suited to calculate the effective $\kappa$ in a large system, since it requires a comparatively smaller computational effort~\cite{Melis2014}. We observe a quantitative difference between the approach described above and AEMD, which is reflected in the extracted value of the thermal conductance of the hBN-hBN interface, $C_{\rm hBN-hBN}=5.27\,{\rm GW}/{\rm m}^2{\rm K}$. From the data reported in Ref.~\cite{Hahn2016}, we also estimate the thermal conductance of the G-G interface to be $C_{\rm G-G}=12.66\,{\rm GW}/{\rm m}^2{\rm K}$. We attribute these lower values to the structure of the GBs investigated; the GBs in the AEMD calculations tend to be disordered and meandering, as shown in Figure~\ref{Samples}, while the GBs used in the NEMD method were mirror symmetric and perfectly periodic arrays of pentagon-heptagon pairs. The inset of Figure~\ref{thermal}(b) shows the thermal conductivity of polycrystalline graphene and hBN using the FE method and the GB conductances extracted from the AEMD method. The smaller values of GB conductivity manifest themselves in a lower overall thermal conductivity, but the main trend holds, and a grain size of 100 nm still appears to be the crossover where thermal transport is no longer dominated by the GBs.

To summarize, we have presented an electrical and thermal characterization of coplanar G-hBN heterostructures. The tight-binding model includes a refined description of the G-hBN interfaces, and is used to describe the electrical properties of polycrystalline structures with varying percentages of graphene and hBN. Our results reproduce the transition from graphene to insulating hBN, with an electrical conductivity change of more than two orders of magnitude and a strong suppression of the Seebeck coefficient. Additionally, the thermal conductivity of these polycrystalline structures has been investigated using a combination of atomistic MD simulations and a FE evaluation of the heat equation. We find that for small-grain structures, the thermal conductivity is minimized for a hBN grain density of 70\%. From our study, we can evaluate the upper value of the thermoelectric figure of merit, $ZT=\sigma S^2T/\kappa$. For example, in the case of 40~nm average grain size and 20\%~hBN, $ZT \sim 1\times 10^{-4}$ for a carrier concentration $n=5\times 10^{12} {\rm cm}^{-2}$, which is quite small. Even for energies near the edge of the gap, where the Seebeck coefficient should be maximized, the value of $ZT$ only reaches $\sim1\times 10^{-2}$.

\begin{acknowledgement}
 J.E.B.-V. acknowledges support from CONACyT (Mexico, D.F.). This work was supported by European Union Seventh Framework Programme under grant agreement 604391 Graphene Flagship (R.M.G.). S. R. acknowledges the Spanish Ministry of Economy and Competitiveness for funding (MAT2012-33911), the Secretaria de Universidades e Investigacion del Departamento de Economia y Conocimiento de la Generalidad de Cataluna and the Severo Ochoa Program (MINECO SEV-2013-0295). M.P. and L.C. acknowledge Spanish MINECO (FIS2015-64886-C5-3-P) and Generalitat de Catalunya (2014SGR301). B.M. and T.R. greatly acknowledge the financial support by European Research Council for COMBAT project (Grant number 615132).
\end{acknowledgement}

\begin{suppinfo}
Details of tight-binding model, details on the numerical evaluation of electrical and thermal conductivity, and scaling analysis to estimate electrical and thermal GB resistivity.
\end{suppinfo}

\bibliography{bibGrhBN}{}

\providecommand{\latin}[1]{#1}
\providecommand*\mcitethebibliography{\thebibliography}
\csname @ifundefined\endcsname{endmcitethebibliography}
  {\let\endmcitethebibliography\endthebibliography}{}
\begin{mcitethebibliography}{39}
\providecommand*\natexlab[1]{#1}
\providecommand*\mciteSetBstSublistMode[1]{}
\providecommand*\mciteSetBstMaxWidthForm[2]{}
\providecommand*\mciteBstWouldAddEndPuncttrue
  {\def\EndOfBibitem{\unskip.}}
\providecommand*\mciteBstWouldAddEndPunctfalse
  {\let\EndOfBibitem\relax}
\providecommand*\mciteSetBstMidEndSepPunct[3]{}
\providecommand*\mciteSetBstSublistLabelBeginEnd[3]{}
\providecommand*\EndOfBibitem{}
\mciteSetBstSublistMode{f}
\mciteSetBstMaxWidthForm{subitem}{(\alph{mcitesubitemcount})}
\mciteSetBstSublistLabelBeginEnd
  {\mcitemaxwidthsubitemform\space}
  {\relax}
  {\relax}

\bibitem[Rubio(2010)]{Rubio2010}
Rubio,~A. \emph{Nat. Mater.} \textbf{2010}, \emph{9}, 379\relax
\mciteBstWouldAddEndPuncttrue
\mciteSetBstMidEndSepPunct{\mcitedefaultmidpunct}
{\mcitedefaultendpunct}{\mcitedefaultseppunct}\relax
\EndOfBibitem
\bibitem[Ci \latin{et~al.}(2010)Ci, Song, Jin, Jariwala, Wu, Li, Srivastava,
  Wang, Storr, Balicas, Liu, and Ajayan]{Ci2010}
Ci,~L.; Song,~L.; Jin,~C.; Jariwala,~D. \latin{et~al.}  \emph{Nat. Mater.}
  \textbf{2010}, \emph{9}, 430\relax
\mciteBstWouldAddEndPuncttrue
\mciteSetBstMidEndSepPunct{\mcitedefaultmidpunct}
{\mcitedefaultendpunct}{\mcitedefaultseppunct}\relax
\EndOfBibitem
\bibitem[Novoselov \latin{et~al.}(2004)Novoselov, Geim, Morozov, Jiang, Zhang,
  Dubonos, Grigorieva, and Firsov]{Novoselov2004}
Novoselov,~K.~S.; Geim,~A.~K.; Morozov,~S.~V.; Jiang,~D. \latin{et~al.}
  \emph{Science} \textbf{2004}, \emph{306}, 666--669\relax
\mciteBstWouldAddEndPuncttrue
\mciteSetBstMidEndSepPunct{\mcitedefaultmidpunct}
{\mcitedefaultendpunct}{\mcitedefaultseppunct}\relax
\EndOfBibitem
\bibitem[Balandin \latin{et~al.}(2008)Balandin, Ghosh, Bao, Calizo,
  Teweldebrhan, Miao, and Lau]{Baladin2008}
Balandin,~A.~A.; Ghosh,~S.; Bao,~W.; Calizo,~I. \latin{et~al.}  \emph{Nano
  Lett.} \textbf{2008}, \emph{8}, 902--907\relax
\mciteBstWouldAddEndPuncttrue
\mciteSetBstMidEndSepPunct{\mcitedefaultmidpunct}
{\mcitedefaultendpunct}{\mcitedefaultseppunct}\relax
\EndOfBibitem
\bibitem[Jo \latin{et~al.}(2013)Jo, Pettes, Kim, Watanabe, Taniguchi, Yao, and
  Shi]{Jo2013}
Jo,~I.; Pettes,~M.~T.; Kim,~J.; Watanabe,~K. \latin{et~al.}  \emph{Nano Lett.}
  \textbf{2013}, \emph{13}, 550--554\relax
\mciteBstWouldAddEndPuncttrue
\mciteSetBstMidEndSepPunct{\mcitedefaultmidpunct}
{\mcitedefaultendpunct}{\mcitedefaultseppunct}\relax
\EndOfBibitem
\bibitem[Wang \latin{et~al.}(2016)Wang, Guo, Dong, Aiyiti, Xu, and
  Li]{Wang2016}
Wang,~C.; Guo,~J.; Dong,~L.; Aiyiti,~A. \latin{et~al.}  \emph{Sci. Rep.}
  \textbf{2016}, \emph{6}, 25334\relax
\mciteBstWouldAddEndPuncttrue
\mciteSetBstMidEndSepPunct{\mcitedefaultmidpunct}
{\mcitedefaultendpunct}{\mcitedefaultseppunct}\relax
\EndOfBibitem
\bibitem[Levendorf \latin{et~al.}(2012)Levendorf, Kim, Brown, Huang, Havener,
  Muller, and Park]{Levendorf2012}
Levendorf,~M.~P.; Kim,~C.-J.; Brown,~L.; Huang,~P.~Y. \latin{et~al.}
  \emph{Nature} \textbf{2012}, \emph{488}, 627--632\relax
\mciteBstWouldAddEndPuncttrue
\mciteSetBstMidEndSepPunct{\mcitedefaultmidpunct}
{\mcitedefaultendpunct}{\mcitedefaultseppunct}\relax
\EndOfBibitem
\bibitem[Liu \latin{et~al.}(2013)Liu, Ma, Shi, Zhou, Gong, Lei, Yang, Zhang,
  Yu, Hackenberg, \latin{et~al.} others]{Liu2013}
Liu,~Z.; Ma,~L.; Shi,~G.; Zhou,~W. \latin{et~al.}  \emph{Nat. Nanotechnol.}
  \textbf{2013}, \emph{8}, 119--124\relax
\mciteBstWouldAddEndPuncttrue
\mciteSetBstMidEndSepPunct{\mcitedefaultmidpunct}
{\mcitedefaultendpunct}{\mcitedefaultseppunct}\relax
\EndOfBibitem
\bibitem[Han \latin{et~al.}(2013)Han, Rodr\'{i}guez-Manzo, Lee, Kybert, Lerner,
  Qi, Dattoli, Rappe, Drndic, and Johnson]{Gang2013}
Han,~G.~H.; Rodr\'{i}guez-Manzo,~J.~A.; Lee,~C.-W.; Kybert,~N.~J.
  \latin{et~al.}  \emph{ACS Nano} \textbf{2013}, \emph{7}, 10129--10138\relax
\mciteBstWouldAddEndPuncttrue
\mciteSetBstMidEndSepPunct{\mcitedefaultmidpunct}
{\mcitedefaultendpunct}{\mcitedefaultseppunct}\relax
\EndOfBibitem
\bibitem[Liu \latin{et~al.}(2014)Liu, Park, Siegel, McCarty, Clark, Deng,
  Basile, Idrobo, Li, and Gu]{Liu2014}
Liu,~L.; Park,~J.; Siegel,~D.~A.; McCarty,~K.~F. \latin{et~al.}  \emph{Science}
  \textbf{2014}, \emph{343}, 163--167\relax
\mciteBstWouldAddEndPuncttrue
\mciteSetBstMidEndSepPunct{\mcitedefaultmidpunct}
{\mcitedefaultendpunct}{\mcitedefaultseppunct}\relax
\EndOfBibitem
\bibitem[Zhao \latin{et~al.}(2012)Zhao, Wang, Yang, Liu, and Liu]{Zhao2012}
Zhao,~R.; Wang,~J.; Yang,~M.; Liu,~Z. \latin{et~al.}  \emph{The Journal of
  Physical Chemistry C} \textbf{2012}, \emph{116}, 21098--21103\relax
\mciteBstWouldAddEndPuncttrue
\mciteSetBstMidEndSepPunct{\mcitedefaultmidpunct}
{\mcitedefaultendpunct}{\mcitedefaultseppunct}\relax
\EndOfBibitem
\bibitem[Matthes \latin{et~al.}(2012)Matthes, Hannewald, and
  Bechstedt]{Lars2012}
Matthes,~L.; Hannewald,~K.; Bechstedt,~F. \emph{Phys. Rev. B} \textbf{2012},
  \emph{86}, 205409\relax
\mciteBstWouldAddEndPuncttrue
\mciteSetBstMidEndSepPunct{\mcitedefaultmidpunct}
{\mcitedefaultendpunct}{\mcitedefaultseppunct}\relax
\EndOfBibitem
\bibitem[Gong \latin{et~al.}(2014)Gong, Shi, Zhang, Zhou, Jung, Gao, Ma, Yang,
  Yang, You, \latin{et~al.} others]{Gong2014}
Gong,~Y.; Shi,~G.; Zhang,~Z.; Zhou,~W. \latin{et~al.}  \emph{Nat. Commun.}
  \textbf{2014}, \emph{5}, 3193\relax
\mciteBstWouldAddEndPuncttrue
\mciteSetBstMidEndSepPunct{\mcitedefaultmidpunct}
{\mcitedefaultendpunct}{\mcitedefaultseppunct}\relax
\EndOfBibitem
\bibitem[Song \latin{et~al.}(2012)Song, Balicas, Mowbray, Capaz, Storr, Ci,
  Jariwala, Kurth, Louie, Rubio, and Ajayan]{Li2012}
Song,~L.; Balicas,~L.; Mowbray,~D.~J.; Capaz,~R.~B. \latin{et~al.}  \emph{Phys.
  Rev. B} \textbf{2012}, \emph{86}, 075429\relax
\mciteBstWouldAddEndPuncttrue
\mciteSetBstMidEndSepPunct{\mcitedefaultmidpunct}
{\mcitedefaultendpunct}{\mcitedefaultseppunct}\relax
\EndOfBibitem
\bibitem[Tuan \latin{et~al.}(2013)Tuan, Kotakoski, Louvet, Ortmann, Meyer, and
  Roche]{Tuan2013}
Tuan,~D.~V.; Kotakoski,~J.; Louvet,~T.; Ortmann,~F. \latin{et~al.}  \emph{Nano
  Lett.} \textbf{2013}, \emph{13}, 1730--1735\relax
\mciteBstWouldAddEndPuncttrue
\mciteSetBstMidEndSepPunct{\mcitedefaultmidpunct}
{\mcitedefaultendpunct}{\mcitedefaultseppunct}\relax
\EndOfBibitem
\bibitem[Cummings \latin{et~al.}(2014)Cummings, Duong, Nguyen, Van~Tuan,
  Kotakoski, Barrios~Vargas, Lee, and Roche]{Cummings2014}
Cummings,~A.~W.; Duong,~D.~L.; Nguyen,~V.~L.; Van~Tuan,~D. \latin{et~al.}
  \emph{Adv. Mater.} \textbf{2014}, \emph{26}, 5079--5094\relax
\mciteBstWouldAddEndPuncttrue
\mciteSetBstMidEndSepPunct{\mcitedefaultmidpunct}
{\mcitedefaultendpunct}{\mcitedefaultseppunct}\relax
\EndOfBibitem
\bibitem[Mortazavi \latin{et~al.}(2014)Mortazavi, Potschke, and
  Cuniberti]{Mortazavi2014}
Mortazavi,~B.; Potschke,~M.; Cuniberti,~G. \emph{Nanoscale} \textbf{2014},
  \emph{6}, 3344--3352\relax
\mciteBstWouldAddEndPuncttrue
\mciteSetBstMidEndSepPunct{\mcitedefaultmidpunct}
{\mcitedefaultendpunct}{\mcitedefaultseppunct}\relax
\EndOfBibitem
\bibitem[Hahn \latin{et~al.}(2016)Hahn, Melis, and Colombo]{Hahn2016}
Hahn,~K.~R.; Melis,~C.; Colombo,~L. \emph{Carbon} \textbf{2016}, \emph{96},
  429--438\relax
\mciteBstWouldAddEndPuncttrue
\mciteSetBstMidEndSepPunct{\mcitedefaultmidpunct}
{\mcitedefaultendpunct}{\mcitedefaultseppunct}\relax
\EndOfBibitem
\bibitem[Isacsson \latin{et~al.}(2017)Isacsson, Cummings, Colombo, Colombo,
  Kinaret, and Roche]{Andreas2017}
Isacsson,~A.; Cummings,~A.~W.; Colombo,~L.; Colombo,~L. \latin{et~al.}
  \emph{2D Materials} \textbf{2017}, \emph{4}, 012002\relax
\mciteBstWouldAddEndPuncttrue
\mciteSetBstMidEndSepPunct{\mcitedefaultmidpunct}
{\mcitedefaultendpunct}{\mcitedefaultseppunct}\relax
\EndOfBibitem
\bibitem[Ma \latin{et~al.}(2014)Ma, Sun, Zhao, Li, Li, Zhao, Wang, Luo, Yang,
  Wang, and Hou]{Ma2014}
Ma,~C.; Sun,~H.; Zhao,~Y.; Li,~B. \latin{et~al.}  \emph{Phys. Rev. Lett.}
  \textbf{2014}, \emph{112}, 226802\relax
\mciteBstWouldAddEndPuncttrue
\mciteSetBstMidEndSepPunct{\mcitedefaultmidpunct}
{\mcitedefaultendpunct}{\mcitedefaultseppunct}\relax
\EndOfBibitem
\bibitem[Tison \latin{et~al.}(2014)Tison, Lagoute, Repain, Chacon, Girard,
  Joucken, Sporken, Gargiulo, Yazyev, and Rousset]{Yann2014}
Tison,~Y.; Lagoute,~J.; Repain,~V.; Chacon,~C. \latin{et~al.}  \emph{Nano
  Lett.} \textbf{2014}, \emph{14}, 6382--6386\relax
\mciteBstWouldAddEndPuncttrue
\mciteSetBstMidEndSepPunct{\mcitedefaultmidpunct}
{\mcitedefaultendpunct}{\mcitedefaultseppunct}\relax
\EndOfBibitem
\bibitem[Luican-Mayer \latin{et~al.}(2016)Luican-Mayer, Barrios-Vargas,
  Falkenberg, Aut\`{e}s, Cummings, Soriano, Li, Brandbyge, Yazyev, Roche, and
  Andrei]{Adina2016}
Luican-Mayer,~A.; Barrios-Vargas,~J.~E.; Falkenberg,~J.~T.; Aut\`{e}s,~G.
  \latin{et~al.}  \emph{2D Mater.} \textbf{2016}, \emph{3}, 031005\relax
\mciteBstWouldAddEndPuncttrue
\mciteSetBstMidEndSepPunct{\mcitedefaultmidpunct}
{\mcitedefaultendpunct}{\mcitedefaultseppunct}\relax
\EndOfBibitem
\bibitem[Li \latin{et~al.}(2015)Li, Zou, Liu, Sun, Gao, Qi, Zhou, Yakobson,
  Zhang, and Liu]{Qiucheng2015}
Li,~Q.; Zou,~X.; Liu,~M.; Sun,~J. \latin{et~al.}  \emph{Nano Lett.}
  \textbf{2015}, \emph{15}, 5804--5810\relax
\mciteBstWouldAddEndPuncttrue
\mciteSetBstMidEndSepPunct{\mcitedefaultmidpunct}
{\mcitedefaultendpunct}{\mcitedefaultseppunct}\relax
\EndOfBibitem
\bibitem[Drost \latin{et~al.}(2014)Drost, Uppstu, Schulz,
  H{\"a}m{\"a}l{\"a}inen, Ervasti, Harju, and Liljeroth]{Robert2014}
Drost,~R.; Uppstu,~A.; Schulz,~F.; H{\"a}m{\"a}l{\"a}inen,~S.~K. \latin{et~al.}
   \emph{Nano Lett.} \textbf{2014}, \emph{14}, 5128--5132\relax
\mciteBstWouldAddEndPuncttrue
\mciteSetBstMidEndSepPunct{\mcitedefaultmidpunct}
{\mcitedefaultendpunct}{\mcitedefaultseppunct}\relax
\EndOfBibitem
\bibitem[Lu \latin{et~al.}(2014)Lu, Gomes, Nunes, Neto, and Loh]{Jiong2014}
Lu,~J.; Gomes,~L.~C.; Nunes,~R.~W.; Neto,~A. H.~C. \latin{et~al.}  \emph{Nano
  Lett.} \textbf{2014}, \emph{14}, 5133--5139\relax
\mciteBstWouldAddEndPuncttrue
\mciteSetBstMidEndSepPunct{\mcitedefaultmidpunct}
{\mcitedefaultendpunct}{\mcitedefaultseppunct}\relax
\EndOfBibitem
\bibitem[Wang \latin{et~al.}(2014)Wang, Song, and Xu]{Wang2014}
Wang,~Y.; Song,~Z.; Xu,~Z. \emph{J. Mater. Res.} \textbf{2014}, \emph{29},
  362--372\relax
\mciteBstWouldAddEndPuncttrue
\mciteSetBstMidEndSepPunct{\mcitedefaultmidpunct}
{\mcitedefaultendpunct}{\mcitedefaultseppunct}\relax
\EndOfBibitem
\bibitem[Liu \latin{et~al.}(2014)Liu, Lin, and Luo]{Liu2014Th}
Liu,~H.~K.; Lin,~Y.; Luo,~S.~N. \emph{J. Phys. Chem. C} \textbf{2014},
  \emph{118}, 24797--24802\relax
\mciteBstWouldAddEndPuncttrue
\mciteSetBstMidEndSepPunct{\mcitedefaultmidpunct}
{\mcitedefaultendpunct}{\mcitedefaultseppunct}\relax
\EndOfBibitem
\bibitem[Mortazavi \latin{et~al.}(2015)Mortazavi, Pereira, Jiang, and
  Rabczuk]{Mortazavi2015}
Mortazavi,~B.; Pereira,~L. F.~C.; Jiang,~J.-W.; Rabczuk,~T. \emph{Sci. Rep.}
  \textbf{2015}, \emph{5}\relax
\mciteBstWouldAddEndPuncttrue
\mciteSetBstMidEndSepPunct{\mcitedefaultmidpunct}
{\mcitedefaultendpunct}{\mcitedefaultseppunct}\relax
\EndOfBibitem
\bibitem[Plimpton(1995)]{Plimpton1995}
Plimpton,~S. \emph{J. Comp. Phys.} \textbf{1995}, \emph{117}, 1--19\relax
\mciteBstWouldAddEndPuncttrue
\mciteSetBstMidEndSepPunct{\mcitedefaultmidpunct}
{\mcitedefaultendpunct}{\mcitedefaultseppunct}\relax
\EndOfBibitem
\bibitem[Brenner \latin{et~al.}(2002)Brenner, Shenderova, Harrison, Stuart, Ni,
  and Sinnott]{Brenner2002}
Brenner,~D.~W.; Shenderova,~O.~A.; Harrison,~J.~A.; Stuart,~S.~J.
  \latin{et~al.}  \emph{J. Phys.: Condens. Matter} \textbf{2002}, \emph{14},
  783\relax
\mciteBstWouldAddEndPuncttrue
\mciteSetBstMidEndSepPunct{\mcitedefaultmidpunct}
{\mcitedefaultendpunct}{\mcitedefaultseppunct}\relax
\EndOfBibitem
\bibitem[Mart\'{i}nez-Gordillo(2014)]{Rafael2014thesis}
Mart\'{i}nez-Gordillo,~R. Atomistic simulations in hybrid C/BN structures.
  Ph.D.\ thesis, Universitat Aut\`{o}noma de Barcelona, 2014; Chapter 5\relax
\mciteBstWouldAddEndPuncttrue
\mciteSetBstMidEndSepPunct{\mcitedefaultmidpunct}
{\mcitedefaultendpunct}{\mcitedefaultseppunct}\relax
\EndOfBibitem
\bibitem[Roche(1999)]{Roche1999}
Roche,~S. \emph{Phys. Rev. B} \textbf{1999}, \emph{59}, 2284--2291\relax
\mciteBstWouldAddEndPuncttrue
\mciteSetBstMidEndSepPunct{\mcitedefaultmidpunct}
{\mcitedefaultendpunct}{\mcitedefaultseppunct}\relax
\EndOfBibitem
\bibitem[Torres \latin{et~al.}(2014)Torres, Roche, and Charlier]{RocheBook}
Torres,~L. E. F.~F.; Roche,~S.; Charlier,~J.-C. \emph{Introduction to
  Graphene-Based Nanomaterials}; Cambridge University Press: Cambridge, UK,
  2014\relax
\mciteBstWouldAddEndPuncttrue
\mciteSetBstMidEndSepPunct{\mcitedefaultmidpunct}
{\mcitedefaultendpunct}{\mcitedefaultseppunct}\relax
\EndOfBibitem
\bibitem[Dorgan \latin{et~al.}(2010)Dorgan, Bae, and Pop]{Dorgan2010}
Dorgan,~V.~E.; Bae,~M.-H.; Pop,~E. \emph{Appl. Phys. Lett.} \textbf{2010},
  \emph{97}, 082112\relax
\mciteBstWouldAddEndPuncttrue
\mciteSetBstMidEndSepPunct{\mcitedefaultmidpunct}
{\mcitedefaultendpunct}{\mcitedefaultseppunct}\relax
\EndOfBibitem
\bibitem[Woessner \latin{et~al.}(2016)Woessner, Alonso-Gonz\'{a}lez, Lundeberg,
  Gao, Barrios-Vargas, Navickaite, Ma, Janner, Watanabe, Cummings, Taniguchi,
  Pruneri, Roche, Jarillo-Herrero, Hone, Hillenbrand, and
  Koppens]{Woessner2016}
Woessner,~A.; Alonso-Gonz\'{a}lez,~P.; Lundeberg,~M.~B.; Gao,~Y. \latin{et~al.}
   \emph{Nat. Commun.} \textbf{2016}, \emph{7}, 10783\relax
\mciteBstWouldAddEndPuncttrue
\mciteSetBstMidEndSepPunct{\mcitedefaultmidpunct}
{\mcitedefaultendpunct}{\mcitedefaultseppunct}\relax
\EndOfBibitem
\bibitem[Sevin\ifmmode~\mbox{\c{c}}\else \c{c}\fi{}li
  \latin{et~al.}(2011)Sevin\ifmmode~\mbox{\c{c}}\else \c{c}\fi{}li, Li, Mingo,
  Cuniberti, and Roche]{Sevincli2011}
Sevin\ifmmode~\mbox{\c{c}}\else \c{c}\fi{}li,~H.; Li,~W.; Mingo,~N.;
  Cuniberti,~G. \latin{et~al.}  \emph{Phys. Rev. B} \textbf{2011}, \emph{84},
  205444\relax
\mciteBstWouldAddEndPuncttrue
\mciteSetBstMidEndSepPunct{\mcitedefaultmidpunct}
{\mcitedefaultendpunct}{\mcitedefaultseppunct}\relax
\EndOfBibitem
\bibitem[K{\i}nac{\i} \latin{et~al.}(2012)K{\i}nac{\i}, Haskins, Sevik, and
  \c{C}a\u{g}{\i}n]{Kinaci2012}
K{\i}nac{\i},~A.; Haskins,~J.~B.; Sevik,~C.; \c{C}a\u{g}{\i}n,~T. \emph{Phys.
  Rev. B} \textbf{2012}, \emph{86}, 115410\relax
\mciteBstWouldAddEndPuncttrue
\mciteSetBstMidEndSepPunct{\mcitedefaultmidpunct}
{\mcitedefaultendpunct}{\mcitedefaultseppunct}\relax
\EndOfBibitem
\bibitem[Melis \latin{et~al.}(2014)Melis, Dettori, Vandermeulen, and
  Colombo]{Melis2014}
Melis,~C.; Dettori,~R.; Vandermeulen,~S.; Colombo,~L. \emph{Eur. Phys. J. B}
  \textbf{2014}, \emph{87}, 96\relax
\mciteBstWouldAddEndPuncttrue
\mciteSetBstMidEndSepPunct{\mcitedefaultmidpunct}
{\mcitedefaultendpunct}{\mcitedefaultseppunct}\relax
\EndOfBibitem
\end{mcitethebibliography}

\end{document}